\documentclass[twocolumn,showpacs,preprintnumbers,amsmath,amssymb]{revtex4}
\usepackage{graphicx}
\usepackage{dcolumn}
\usepackage{bm}

\usepackage{color}

\usepackage[english]{babel}

\begin{document}
\title{Analytical properties of a three-compartmental dynamical demographic model}

\author{E.B. Postnikov}%
 \email{postnicov@gmail.com}
\affiliation{Department of Theoretical Physics,  Kursk State University,
Radishcheva st., 33, 305000, Kursk, Russia}

\date{2015, July 27}

\begin{abstract}
The three-compartmental demographic model by Korotaeyv-Malkov-Khaltourina, connecting population size, economic surplus, and  educational level, is considered from the point of view of dynamical systems theory. It is shown that there exist two integrals of motion, which enable the system to be reduced to one non-linear ordinary differential equation. The study of its structure provides analytical criteria for the dominance ranges   of the dynamics of Malthus and Kremer. Additionally, the particular ranges of parameters enable the derived general ordinary differential equations to be reduced to the models of Gompertz and Thoularis-Wallace.
\end{abstract}

\maketitle

\section{Introduction}

The Population Division at the United Nations declared that ``Timely and accurate information about population trends continues to be in high demand'' since it ``is needed for the formulation and implementation of policies and programmes in almost all areas of public life'' \cite{UN}. Correspondingly, the mathematical modeling of demographic processes dates back to the work of Malthus, Verhulst, Pearl, and Reed, and it remains a challenge for non-linear dynamics aimed at reproducing and explaining modern population data and predicting future trends \cite{Bloom2011,Lee2011}. 

Two of the most notable features of global demography are the hyperbolic growth of the global population \cite{vonFoerster1960} 
\begin{equation}
N=\frac{N_0}{1-bN_0t};\,b,\,N_0=\mathrm{const},
\label{hyperN}
\end{equation}
 which is traced from 1960--1970th back to prehistoric times \cite{vonFoerster1960,Kremer1993}, and the global demographic transition, which slows down this population growth and tends to stabilize the overall number of people \cite{Galor2000,LutzBook2013}.

The hyperbolic function (\ref{hyperN}) is a solution of the equation  
\begin{equation}
\frac{dN}{dt}=bN^2,
\label{hypODE}
\end{equation}
which was proposed in \cite{vonFoerster1960} considering the Malthusian dynamics
\begin{equation}
\frac{dN}{dt}=aN,
\label{Malth}
\end{equation}
 with the population-dependent growth rate $a=bN$, which was hypothesized as an effect of people's cooperativity. 
A more precise background of this effect was revealed by Kremer \cite{Kremer1993}, who argued that the functional form (\ref{hypODE}) originates from the Malthusian economic growth and the fast stabilizing balance between actual economic levels and population sizes. 

On the other hand, the demographic transition is a sufficiently more complex process. Its description requires taking into account several additional social factors leading to a  saturation of the population, and, correspondingly, more sophisticated models of non-linear dynamics. From the latter point of view, Ref.~\cite{Cabella2011} proved the existence of the unified scaling approach not only to  the generalized model of von Foerster {\it et al.} [$a(N)=bN^{\alpha}$ with arbitrary $\alpha$] but also to the Thoularis-Wallace model \cite{Tsoularis2002} of generalized logistic growth
\begin{equation}
\frac{dN}{dt}=bN^{\alpha}\left[1-\left(\frac{N}{K}\right)^{\beta}\right]^{\gamma},
\label{TSW}
\end{equation}
where $K$ is the carrying capacity and $\alpha$, $\beta$, and $\gamma$ are the adjustable parameters.
However, the developed mathematical framework leaves  open the question of the origin of non-integer values of these parameters. 

Some recent sociological studies \cite{Lutz2011,Murtin2013,Warren2015} argue that education level is one of the key factors that effectively stops world population growth. The primary effect is a lower overall fertility rate for well-educated women, which could dominate other economic and social factors.

At the same time, Korotayev, Malkov and Khaltourina proposed \cite{Korotaev2006book} a relatively simple model that connects these three factors in a natural way:
\begin{eqnarray}
\frac{dN}{dt}&=&aNS(1-L),\label{eqN}\\
\frac{dS}{dt}&=&bNS,\label{eqS}\\
\frac{dL}{dt}&=&cSL(1-L).\label{eqL}
\end{eqnarray}
Here the population growth (\ref{eqN}) satisfies Malthusian dynamics with the growth's rate supported by economic surplus $S$ and restricted by education level $L$ (the authors of \cite{Korotaev2006book} considered literacy to be such an indicator; however, some other educational determinants could be considered without loss of generality). The economic development (\ref{eqS}) admits Kremer's assumption, and education level satisfies the Verhulst form, in which the growth rate is determined by the surplus. The coefficients $a$, $b$, and $c$ are constants. 

It has been shown \cite{Korotaev2006book} that numerical solutions of the system (\ref{eqN})--(\ref{eqL}) correspond to global statistical data with sufficiently high accuracy. However, all of these conclusions, including the detection of an interval, where $N$ mimics hyperbolic growth (\ref{hyperN}) were the results of numerical simulations only. 

Thus, the main goal of the present work is to explore Eqs.~(\ref{eqN})--(\ref{eqL}) analytically as a non-linear dynamical system. The principal questions are as follows: (i) What is its actual dimension, i.e., the existence of integrals of motion, that determines the degrees of freedom for an adjustment to the demographic data? (ii) Is it really reducible to the classical models of demography and population dynamics in general, and what are the corresponding ranges of population size? 

\section{Results}

\subsection{Basic reduction of ordinary differential equations}

Since the system (\ref{eqN})--(\ref{eqL}) contains similar products of the same variables, it is natural to examine scaling functions and integrals of motion considering derivatives of $N$, $S$, and $L$ with respect to each other, excluding time. 

It should be pointed out that one integral was found directly by the authors of the model dividing Eq.~(\ref{eqN}) by Eq.~(\ref{eqL}) that results in the power law dependence  $L=\lambda N^{\frac{c}{a}}$, where the factor $\lambda=L_0N_0^{-\frac{c}{a}}$ is determined by the initial conditions.

At the same time, this procedure can be successfully continued, i.e., the system (\ref{eqN})--(\ref{eqL}) can be reduced to one equation in an explicit form. Substitution of the revealed function $L=L(N)$ into Eq.~(\ref{eqN}),
\begin{eqnarray}
\frac{dN}{dt}&=&aNS\left(1-\lambda N^{\frac{c}{a}}\right),\label{eqN1}\\
\frac{dS}{dt}&=&bNS,\label{eqS1}
\end{eqnarray}
and dividing Eq.~(\ref{eqN1}) by Eq.~(\ref{eqS1}) leads to the equation
$$
\frac{dN}{dS}=\frac{a}{b}\left(1-\lambda N^{\frac{c}{a}}\right),
$$
which allows for the separation of variables,
$$
\int_{N_0}^{N}\frac{dN}{1-\lambda N^{\frac{c}{a}}}=\frac{a}{b}\left(S-S_0\right).
$$
The integral on the left-hand side has a closed-form expression via the Gauss hypergeometric function $_2F_1$ \cite{Abramowitz1964handbook}. Thus, the solution explicitly representing $N$ as a function of $S$, is
\begin{equation}
S=\frac{b}{a}N{}_2F_1\left(1,\frac{a}{c};1+\frac{a}{c};\lambda N^{\frac{c}{a}}\right)+\mu(N_0,S_0),
\label{SN}
\end{equation}
where 
\begin{equation}
\mu(N_0,S_0)=S_0-\frac{b}{a}N{}_2F_1\left(1,\frac{a}{c};1+\frac{a}{c};\lambda N_0^{\frac{c}{a}}\right).
\label{mu0}
\end{equation}

Therefore, both $L$ and $S$ can be substituted into Eq.~(\ref{eqN}) as functions of population number, and the system Eq.~(\ref{eqN}) by Eq.~(\ref{eqL}) will be reduced to the unique equation with respect to this variable:
\begin{eqnarray}
\frac{dN}{dt}&=&bN^2{}_2F_1\left(1,\frac{a}{c};1+\frac{a}{c};\lambda N^{\frac{c}{a}}\right)\left(1-\lambda N^{\frac{c}{a}}\right) \nonumber \\&&+ab\mu N\left(1-\lambda N^{\frac{c}{a}}\right).
\label{eqNfull0}
\end{eqnarray}
It should be supplied with an initial value $N(0)=N_0$.

The hypergeometric function is given by the Gauss series,
$$
{}_2F_1\left(1,\frac{a}{c};1+\frac{a}{c};\lambda N^{\frac{c}{a}}\right)=
\frac{\Gamma\left(1+\frac{a}{c}\right)}{\Gamma\left(\frac{a}{c}\right)}
\sum\limits_{n=0}^{\infty}\frac{\Gamma\left(\frac{a}{c}+n\right)(\lambda N^{\frac{c}{a}})^n }{\Gamma\left(1+\frac{a}{c}+n\right)},
$$
which can be simplified using the functional relation $\Gamma(1+z)=z\Gamma(z)$. 

As a result, Eq.~(\ref{eqNfull0}) takes the form
\begin{equation}
\frac{dN}{dt}=\left[
ab\mu N+bN^2+bN^2\sum\limits_{n=1}^{\infty}\frac{\left(\lambda N^{\frac{c}{a}}\right)^n}{\left(1+n\frac{c}{a}\right)}
\right]\left(1-\lambda N^{\frac{c}{a}}\right),
\label{eqNfull}
\end{equation}
which is quite demonstrable for a qualitative analysis.

The second multiplier provides a delimiter, which determines the stable maximal population's size $N_s=\lambda^{-1/(c/a)}$ corresponding to the education level $L=1$. 

The first multiplier consists of three terms. Each one corresponds to the basic demographic process: (i) $ab\mu N$ is the simplest Malthusian growth; (ii) $bN^2$ satisfies Kremer's economic-demographic model; and (iii) the last series describes the influence of education level on the rate of population growth.
Each term dominates within its own range of $N$.

\subsection{Discussion of particular cases}

Small populations with an education level far from saturation ($N<<a\mu$ and $N<<\lambda^{-1/(c/a)}$) grow exponentially as $N\propto\exp(ab\mu t)$ since Eq.~(\ref{eqNfull}) reduces to the simplest Malthus law, $dN/dt=ab\mu N$.

The transition from exponential  to hyperbolic growth occurs for $N>a\mu$, but still $N<<\lambda^{-1/(c/a)}$. The complex interplay of  education level and population size is valuable at the area of the demographic transition only, when $\lambda N^{c/a}$ tends to unity. 

Note also that the global dynamics of the world's population exhibited purely hyperbolic behavior \cite{vonFoerster1960,Kremer1993} for the extremely long time interval before the end of the 20th century. This kind of solution is provided by Eq.~(\ref{eqNfull}) for $\lambda N^{c/a}$ if $\mu=0$. The last equality corresponds to the quite notable invariance property of the Cauchy problem stated for the full system (\ref{eqN})--(\ref{eqL}). This condition means that the equality
\begin{equation}
S=\frac{b}{a}N{}_2F_1\left(1,\frac{a}{c};1+\frac{a}{c};L\right)
\label{invSNL}
\end{equation}
holds for all moments of time, from the initial one up to an arbitrary one. This follows from (\ref{SN}) and (\ref{mu0}) with the substituted value $\mu=0$. 

The invariant equality (\ref{invSNL}) corresponds explicitly  to Kremer's basic assumption that the population adjusts instantaneously to an economic indicator, but with the additional factor depending on education level $L$. This factor varies from ${}_2F_1\left(1,\frac{a}{c};1+\frac{a}{c};0\right)=1$ (and it is close to $1$ for $L<<1$)  up to ${}_2F_1\left(1,\frac{a}{c};1+\frac{a}{c};1\right)=a/c$.

Thus, the asymptotic balance between economic indicator $S$ and  population size $N$ is represented by the values of the invariant (\ref{invSNL}) as $S=ba^{-1}N$ for the low-educated population ($L=0$) and $S=c^{-1}N$ for the high-educated population ($L=1$). Since the coefficient of proportionality is given as a ratio of  economics and population growth rates $b/a$ in the former case and as the inverse knowledge growth rate $c^{-1}$ only in the latter, the formula (\ref{invSNL}) qualitatively expresses the typical characteristics of the transition from an industrial society to a post-industrial one \cite{Masuda1980book,Mcnicoll2013}. In addition, this effect, in conjunction with the valuable deviation of the Gauss hypergeometrical function from unity, should be taken into account in the solution of Eq.~(\ref{eqNfull0})), when $N$ tends to $\lambda^{-1}$. Thus, its behavior encompasses the coexisting transitions in the demographic curve and in the socio-economical characteristics. 

Consider now some particular cases of parameters that reduce Eq.~(\ref{eqNfull0}) (or, equivalently, Eq.~(\ref{eqNfull})) to known equations of population dynamics. 

The Gauss hypergeometric function has the following expression in elementary functions for $a/c=1$ \cite{Abramowitz1964handbook}:
$$
{}_2F_1\left(1,1;2;\lambda N\right)=-\frac{1}{\lambda N}\log\left(1-\lambda N\right).
$$
Therefore, Eq.~(\ref{eqNfull}) written with respect to dimensionless $n=1-\lambda N$, takes the form
\begin{equation}
\frac{dn}{dt}=bn\left[\log[\exp(a\mu)]+\log(n)\right]\frac{1-n}{\lambda}.
\label{GompG}
\end{equation}
Such an equation (the Gompertz equation multiplied by the Verhulst factor) was considered \cite{Medeiros2001} in the context of the generalization of the Heumann-H{\"o}tzel model, which allows for a reproduction of the drosophila fly population. 

For $n<<1$, Eq.~(\ref{GompG}) reduces to the original Gompertz model
\begin{equation}
\frac{d\log(n)}{dt}=\frac{b}{\lambda}\left[\log[\exp(a\mu)]+\log(n)\right]
\label{Gomp}
\end{equation}
with the solution
\begin{equation}
n\propto e^{\displaystyle{-ke^{\frac{b}{\lambda}t}}},
\label{GompSol}
\end{equation}
where $k$ is a constant depending on $\alpha\mu$ and initial conditions.

Since it was proposed to discuss mortality rate \cite{Gompertz1825,Winsor1932}, the description of the decay of the ``illiterate'' part of a population $n$ is a natural association between models. In addition, the solution (\ref{GompSol}) demonstrates analytically a faster convergence to a stable stationary state in comparison with the logistic dynamics. This fact was mentioned as a result of numerical simulations for $a/c=1.32$ in \cite{Korotaev2006book}, but it did not lead to any explanations or approximations there. Although the solution for $a/c=1.32$  differs quantitatively from the approximation $a/c=1$ (\ref{GompSol}) in a precise value, the principal qualitative behavior originates from the same hyperexponential dynamics. 

Another reduction to a known model can be evaluated for Eq.~(\ref{eqNfull}) taking $a/c<<1$ and $\mu=0$. One can neglect by the sum within the brackets there (in other words, to take ${}_2F_1\left(1,\frac{a}{c};1+\frac{a}{c};\lambda N^{\frac{c}{a}}\right)\approx1$). Obviously, the remaining terms coincide with the Thoularis-Wallace model (\ref{TSW}) with $\alpha=2$, $\beta=c/a$, and $\gamma=1$. Thus, the question about the origin of an empiric fractional Verhulst factor \cite{Tsoularis2002} can be reduced to  consideration of growth rates in the extended system of ordinary differential equations for factors determining  population growth.

\section{Conclusion}

The demographic dynamics, which satisfies Eqs.~(\ref{eqN})--(\ref{eqL}), is one-dimensional and can be reproduced by the single autonomous ordinary differential equation (\ref{eqNfull}). Its coefficient can be taken as adjustable parameters, which can be determined from the known population trends. Toward that end, it is shown that the derived equation (\ref{eqNfull}) reduces to the Malthus model or to the model of von~Foerster~{\it et al.} and Kremer when the population is far from the saturated state. 

The elimination of the Malthusian term is revealed as being directly connected with Kremer's arguments on the balance between economic and  population indicators with the addition of the people's educational factor. The analysis of the latter allows for a demonstration of the transition from an industrial to a post-industrial society in terms of mathematical modeling. 

On the other hand, the interplay of these parameters with the growth rates of the original models and initial conditions places the described approach into the context of the methods of non-linear dynamics, which operate with the decomposition of complex differential equations of physico-chemical kinetics into a set of elementary polynomial ODEs \cite{Wilhelm2000}.

In addition, the particular choice of parameters shows that the form Eq.~(\ref{eqNfull}) is reducible to a variety of known non-trivial models of population dynamics, e.g., the models of Gompertz and Thoularis-Wallace. Thus, it has potential applications beyond just the problems of demography.


\begin{thebibliography}{20}
\expandafter\ifx\csname natexlab\endcsname\relax\def\natexlab#1{#1}\fi
\expandafter\ifx\csname bibnamefont\endcsname\relax
  \def\bibnamefont#1{#1}\fi
\expandafter\ifx\csname bibfnamefont\endcsname\relax
  \def\bibfnamefont#1{#1}\fi
\expandafter\ifx\csname citenamefont\endcsname\relax
  \def\citenamefont#1{#1}\fi
\expandafter\ifx\csname url\endcsname\relax
  \def\url#1{\texttt{#1}}\fi
\expandafter\ifx\csname urlprefix\endcsname\relax\def\urlprefix{URL }\fi
\providecommand{\bibinfo}[2]{#2}
\providecommand{\eprint}[2][]{\url{#2}}

\bibitem[{UN()}]{UN}
\emph{\bibinfo{title}{{United Nations - Department of Economic and Social
  Affairs - Population Division}}},
  \urlprefix\url{http://www.un.org/en/development/desa/population/theme/trends%
/}.

\bibitem[{\citenamefont{Bloom}(2011)}]{Bloom2011}
\bibinfo{author}{\bibfnamefont{D.~E.} \bibnamefont{Bloom}},
  \bibinfo{journal}{Science} \textbf{\bibinfo{volume}{333}},
  \bibinfo{pages}{562} (\bibinfo{year}{2011}).

\bibitem[{\citenamefont{Lee}(2011)}]{Lee2011}
\bibinfo{author}{\bibfnamefont{R.}~\bibnamefont{Lee}},
  \bibinfo{journal}{Science} \textbf{\bibinfo{volume}{333}},
  \bibinfo{pages}{569} (\bibinfo{year}{2011}).

\bibitem[{\citenamefont{von Foerster et~al.}(1960)\citenamefont{von Foerster,
  Mora, and Amiot}}]{vonFoerster1960}
\bibinfo{author}{\bibfnamefont{H.}~\bibnamefont{von Foerster}},
  \bibinfo{author}{\bibfnamefont{P.~M.} \bibnamefont{Mora}}, \bibnamefont{and}
  \bibinfo{author}{\bibfnamefont{L.~W.} \bibnamefont{Amiot}},
  \bibinfo{journal}{Science} \textbf{\bibinfo{volume}{132}},
  \bibinfo{pages}{1291} (\bibinfo{year}{1960}).

\bibitem[{\citenamefont{Kremer}(1993)}]{Kremer1993}
\bibinfo{author}{\bibfnamefont{M.}~\bibnamefont{Kremer}}, \bibinfo{journal}{Q.
  J. Econ.} \textbf{\bibinfo{volume}{108}}, \bibinfo{pages}{681}
  (\bibinfo{year}{1993}).

\bibitem[{\citenamefont{Galor and Weil}(2000)}]{Galor2000}
\bibinfo{author}{\bibfnamefont{O.}~\bibnamefont{Galor}} \bibnamefont{and}
  \bibinfo{author}{\bibfnamefont{D.~N.} \bibnamefont{Weil}},
  \bibinfo{journal}{Am. Econ. Rev.} \textbf{\bibinfo{volume}{90}},
  \bibinfo{pages}{806} (\bibinfo{year}{2000}).

\bibitem[{\citenamefont{Lutz et~al.}(2013)\citenamefont{Lutz, Sanderson, and
  Sherbov}}]{LutzBook2013}
\bibinfo{editor}{\bibfnamefont{W.}~\bibnamefont{Lutz}},
  \bibinfo{editor}{\bibfnamefont{W.~C.} \bibnamefont{Sanderson}},
  \bibnamefont{and} \bibinfo{editor}{\bibfnamefont{S.}~\bibnamefont{Sherbov}},
  eds., \emph{\bibinfo{title}{{The End of World Population Growth in the 21st
  Century: New Challenges for Human Capital Formation and Sustainable
  Development}}} (\bibinfo{publisher}{Routledge, New York},
  \bibinfo{year}{2013}).

\bibitem[{\citenamefont{Cabella et~al.}(2011)\citenamefont{Cabella, Martinez,
  and Ribeiro}}]{Cabella2011}
\bibinfo{author}{\bibfnamefont{B.~C.~T.} \bibnamefont{Cabella}},
  \bibinfo{author}{\bibfnamefont{A.~S.} \bibnamefont{Martinez}},
  \bibnamefont{and} \bibinfo{author}{\bibfnamefont{F.}~\bibnamefont{Ribeiro}},
  \bibinfo{journal}{Phys. Rev. E} \textbf{\bibinfo{volume}{83}},
  \bibinfo{pages}{061902} (\bibinfo{year}{2011}).

\bibitem[{\citenamefont{Tsoularis and Wallace}(2002)}]{Tsoularis2002}
\bibinfo{author}{\bibfnamefont{A.}~\bibnamefont{Tsoularis}} \bibnamefont{and}
  \bibinfo{author}{\bibfnamefont{J.}~\bibnamefont{Wallace}},
  \bibinfo{journal}{Math. Biosci.} \textbf{\bibinfo{volume}{179}},
  \bibinfo{pages}{21} (\bibinfo{year}{2002}).

\bibitem[{\citenamefont{Lutz and Samir}(2011)}]{Lutz2011}
\bibinfo{author}{\bibfnamefont{W.}~\bibnamefont{Lutz}} \bibnamefont{and}
  \bibinfo{author}{\bibfnamefont{K.~C.} \bibnamefont{Samir}},
  \bibinfo{journal}{Science} \textbf{\bibinfo{volume}{333}},
  \bibinfo{pages}{587} (\bibinfo{year}{2011}).

\bibitem[{\citenamefont{Murtin}(2013)}]{Murtin2013}
\bibinfo{author}{\bibfnamefont{F.}~\bibnamefont{Murtin}},
  \bibinfo{journal}{Rev. Econ. Stat.} \textbf{\bibinfo{volume}{95}},
  \bibinfo{pages}{617} (\bibinfo{year}{2013}).

\bibitem[{\citenamefont{Warren}(2015)}]{Warren2015}
\bibinfo{author}{\bibfnamefont{S.~G.} \bibnamefont{Warren}},
  \bibinfo{journal}{Earth's Future} \textbf{\bibinfo{volume}{3}},
  \bibinfo{pages}{82} (\bibinfo{year}{2015}).

\bibitem[{\citenamefont{Korotaev et~al.}(2006)\citenamefont{Korotaev, Malkov,
  and Khaltourina}}]{Korotaev2006book}
\bibinfo{author}{\bibfnamefont{A.~V.} \bibnamefont{Korotaev}},
  \bibinfo{author}{\bibfnamefont{A.}~\bibnamefont{Malkov}}, \bibnamefont{and}
  \bibinfo{author}{\bibfnamefont{D.}~\bibnamefont{Khaltourina}},
  \emph{\bibinfo{title}{Introduction to social macrodynamics: compact
  macromodels of the world system growth}} (\bibinfo{publisher}{Editorial URSS,
  Moscow}, \bibinfo{year}{2006}).

\bibitem[{\citenamefont{Abramowitz and Stegun}(1964)}]{Abramowitz1964handbook}
\bibinfo{author}{\bibfnamefont{M.}~\bibnamefont{Abramowitz}} \bibnamefont{and}
  \bibinfo{author}{\bibfnamefont{I.~A.} \bibnamefont{Stegun}},
  \bibinfo{journal}{Washington: National Bureau of Standards}
  (\bibinfo{year}{1964}).

\bibitem[{\citenamefont{Masuda}(1980)}]{Masuda1980book}
\bibinfo{author}{\bibfnamefont{Y.}~\bibnamefont{Masuda}},
  \emph{\bibinfo{title}{The information society as post-industrial society}}
  (\bibinfo{publisher}{World Future Society, Bethesda}, \bibinfo{year}{1980}).

\bibitem[{\citenamefont{McNicoll}(2013)}]{Mcnicoll2013}
\bibinfo{author}{\bibfnamefont{G.}~\bibnamefont{McNicoll}},
  \bibinfo{journal}{Popul. Dev. Rev.} \textbf{\bibinfo{volume}{38}},
  \bibinfo{pages}{3} (\bibinfo{year}{2013}).

\bibitem[{\citenamefont{de~Medeiros and Onody}(2001)}]{Medeiros2001}
\bibinfo{author}{\bibfnamefont{N.~G.~F.} \bibnamefont{de~Medeiros}}
  \bibnamefont{and} \bibinfo{author}{\bibfnamefont{R.~N.} \bibnamefont{Onody}},
  \bibinfo{journal}{Phys. Rev. E} \textbf{\bibinfo{volume}{64}},
  \bibinfo{pages}{041915} (\bibinfo{year}{2001}).

\bibitem[{\citenamefont{Gompertz}(1825)}]{Gompertz1825}
\bibinfo{author}{\bibfnamefont{B.}~\bibnamefont{Gompertz}},
  \bibinfo{journal}{Philos. Trans. R. Soc. London}
  \textbf{\bibinfo{volume}{115}}, \bibinfo{pages}{513} (\bibinfo{year}{1825}).

\bibitem[{\citenamefont{Winsor}(1932)}]{Winsor1932}
\bibinfo{author}{\bibfnamefont{C.~P.} \bibnamefont{Winsor}},
  \bibinfo{journal}{Proc. Natl. Acad. Sci. USA} \textbf{\bibinfo{volume}{18}},
  \bibinfo{pages}{1} (\bibinfo{year}{1932}).

\bibitem[{\citenamefont{Wilhelm}(2000)}]{Wilhelm2000}
\bibinfo{author}{\bibfnamefont{T.}~\bibnamefont{Wilhelm}}, \bibinfo{journal}{J.
  Math. Chem.} \textbf{\bibinfo{volume}{27}}, \bibinfo{pages}{71}
  (\bibinfo{year}{2000}).

\end{thebibliography}
\end{document}